# "WE'LL FIX IT LATER": EDUCATION, AI, AND THE DEFERRAL OF STUDENT PRIVACY IN EDTECH

**Meghna Manoj Nair, Rachel Greenstadt**

New York University (United States)

**Abstract**

Educational technology (EdTech) platforms collect highly sensitive student data, including behavioral logs, disability records, and academic histories. However, privacy considerations are often postponed rather than treated as a foundational design requirement. We present a mixed-methods study combining 12 semi-structured interviews with EdTech professionals and a privacy policy audit of 48 platforms coded across five dimensions, with strong inter-rater reliability (mean Cohen's $\kappa = 0.781$). Our interviews reveal a recurring organizational pattern in which privacy is recognized as important but deferred across the product lifecycle as organizations prioritize product functionality, growth, funding, and immediate educational outcomes. Responsibility is often delegated to cloud providers, policy documents, or downstream institutions, while limited privacy-related feedback gives organizations little pressure to change these practices. The policy analysis reflects these patterns: platforms describe what data they collect relatively well but provide substantially less information about how that data is subsequently governed. Thirty-three percent make no meaningful Artificial Intelligence (AI) disclosure despite visible AI features, and 73% provide only generic accountability and breach-response language. K–12 platforms perform better on children's consent where regulation creates explicit requirements, but this advantage does not extend to AI governance or accountability. These findings suggest that meaningful improvement requires enforceable institutional and regulatory mechanisms rather than voluntary privacy commitments alone.

## 1 INTRODUCTION

Educational technology (EdTech) is now deeply embedded in schooling. Learning management systems, adaptive tutoring, assessment platforms, and AI-enabled tools collect names, device identifiers, behavioral logs, grades, and, in K-12 settings, disability and Individualized Education Program (IEP) records [1], [2]. These data support personalization and analytics, but they also create a high-stakes privacy environment in which children often cannot negotiate data practices or meaningfully opt out [3], [4]. The consequences of weak governance are concrete. Unacademy exposed more than 20 million user profiles in 2020 [5]; Illuminate Education exposed sensitive student records affecting more than 800,000 New York City students [6]; and the 2024 PowerSchool incident affected tens of millions of records [7]. At the same time, EdTech organizations operate across uneven regulatory environments, including the Children's Online Privacy Protection Act (COPPA), Family Educational Rights and Privacy Act (FERPA), General Data Protection Regulation (GDPR), and India's Digital Personal Data Protection Act (DPDPA) [8]-[12].

Prior work has documented vague policies, excessive data collection, uneven procurement, and organizational underinvestment [9], [10], [13], [14], [17], [19], [23], [24], [27]–[29]. What remains less visible is how privacy decisions are made inside EdTech organizations and why known problems persist. Unlike general software-lifecycle work, we examine these decisions in educational settings where institutions and, for minors, parents or guardians often act as proxy decision-makers, while students may have little ability to choose or leave the technology [15], [16]. We therefore ask: (RQ1) how privacy and security are integrated over the product lifecycle; (RQ2) how responsibility is assigned across organizations and infrastructure; (RQ3) how feedback and the visibility of harm shape prioritization; and (RQ4) how public-facing policies reflect these internal patterns.

We combine 12 semi-structured interviews with a policy audit of 48 platforms. Our central finding is a cycle of privacy deferral: privacy is recognized but postponed, delegation makes the delay appear managed, weak feedback generates little pressure to change, and AI makes downstream data use harder to inspect. The contribution is not that EdTech uniquely neglects privacy, but that mandatory adoption, limited student power, diffuse responsibility, and slow-moving harms weaken the corrective mechanisms available in ordinary consumer markets.

## 2 RELATED WORK

EdTech privacy research has established that student data practices are extensive, difficult to contest, and shaped by institutional dependence. Data-intensive personalization can enable long-term sorting and tracking [1], while continuous collection becomes normalized in EdTech discourse [2]. Work on children's data governance shows fragmented authority over education data [3], and research with EdTech firms identifies persistent ethical challenges around AI adoption [4]. Mandatory adoption makes these concerns especially consequential because students often cannot choose or leave institutionally required platforms [15], [16].

Research on schools and procurement shows that privacy risk is not only a technical problem. K-12 institutions face limited vendor accountability, resource shortages, and uneven security capacity [17], while educators often adopt tools with limited structured guidance [18]. Procurement can rely on trust because institutions cannot independently verify vendor claims [19], and higher-education adoption is frequently driven by convenience and institutional momentum [20]. At the same time, EdTech platforms increasingly transform student interactions into proprietary data assets [21], while public concern is fragmented across surveillance, ethics, and technical security [22]. Further, policy and compliance studies document a gap between disclosure and operational protection. Systematic reviews find broad collection of behavioral, emotional, and social data without consistently meaningful consent [10]. Audits show that public-facing privacy commitments can be incomplete or difficult for schools to evaluate [9], [23], while COPPA remains structurally limited for modern EdTech procurement [8]. Stakeholders may recognize privacy concerns without understanding downstream data practices [24]. Cloud architectures further distribute responsibility across vendors and institutions [25], and privacy-by-design obligations remain difficult to operationalize when provider incentives do not reward early investment [26], [27].

Organizational research helps explain why these failures persist despite awareness. Software teams often treat privacy as a later-stage concern [13], and developers of child-directed apps may approach compliance as a threshold rather than a foundation [14]. Privacy engineering requires sustained organizational commitment [28], while security underinvestment can be a rational response to misaligned incentives [29]. Together, this literature explains important pieces of the problem, but it rarely connects internal EdTech decision-making with the public policies through which organizations represent those decisions. Our study takes a step towards bridging that gap. We examine how privacy is deferred, delegated, and made less visible inside EdTech organizations, then test whether those patterns appear in a 48-platform policy audit. We treat AI as an amplifier rather than the origin of these governance problems: AI increases opacity and third-party data flows, but it enters organizational environments where privacy ownership, feedback, and incentives are already weak.

## 3 METHODOLOGY

We used a mixed-methods design in which interviews explain how and why privacy is deferred, while the policy audit tests whether those dynamics are visible in public commitments. The study received institutional ethics approval; participants gave informed consent and identifying details were removed.

### 3.1 Interviews and sample

We recruited participants through professional networks and LinkedIn, targeting people with direct experience in EdTech development, research, data governance, or institutional deployment. Of approximately 100 people contacted, 15 expressed interest and 12 completed 30-45 minute Zoom interviews. Three additional practitioners were unable to participate because of organizational NDA or compliance restrictions. Participants spanned US and Indian startups, nonprofits, research organizations, and public K-12 districts. We sought conceptual variation rather than statistical representativeness. Participants closer to implementation, including interns and classroom teachers, were included because they had direct experience with how organizational privacy decisions were enacted in day-to-day product or institutional practice; however, their roles may provide less visibility into strategic decision-making. Interviews followed a semi-structured protocol covering privacy prioritization across the product lifecycle, data handling, organizational constraints, responses to incidents, and emerging challenges from AI adoption. AI was included because AI-enabled personalization and third-party model use can introduce additional data flows and make downstream uses of student information less visible. When participants were uncomfortable being recorded, detailed notes were used alongside transcripts.

Table 1. Participant information (anonymized).

| ID | Organization | Role |
|---|---|---|
| P1 | National digital learning nonprofit (USA) | Research project manager |
| P2 | University AI & education research lab (USA) | Research management |
| P3 | Public K-12 school district (USA) | Teacher |
| P4 | Early-stage EdTech startup (USA) | Software engineering intern |
| P5 | Public K-12 school district (USA) | Director of ICT services |
| P6 | Education nonprofit (India) | Deputy general manager |
| P7 | Public K-12 school district (USA) | Technology integration specialist |
| P8 | Large nonprofit EdTech org (USA) | Principal data scientist |

| ID | Organization | Role |
|---|---|---|
| P9 | EdTech firm (India) | Tutor, CS and mathematics |
| P10 | Former EdTech startup (USA) | Co-founder |
| P11 | Adult learning & workforce org (USA) | Product manager / data engineer |
| P12 | Career navigation EdTech startup (USA) | Software development intern |

### 3.2 Qualitative analysis

Analysis was informed by Charmaz's grounded theory principles [30]. The first author conducted line-by-line coding across all 12 interviews, generating more than 80 initial codes, then grouped them into eight focused categories based on recurrence and explanatory weight. The authors met throughout analysis to discuss interpretations. We do not claim analytical saturation; given Non Disclosure Agreements (NDA) and compliance barriers, we instead ground credibility in convergence across organizational roles and contexts, consistent with methodological cautions about using saturation as a universal sample-size criterion [30].

Focused coding consolidated more than 80 line-by-line codes into eight higher-order categories: privacy deferral (RQ1), delegating responsibility (RQ2), weak privacy feedback loops (RQ3), structural inequality in privacy capacity (RQ1-RQ2), AI as a governance frontier (RQ1, RQ3), procedural gatekeeping (RQ2), commercialization as privacy pressure (RQ2, RQ3), and regulatory ambiguity (RQ2). These categories capture recurring mechanisms across participants rather than prevalence estimates: postponing security until after launch, treating compliance as a threshold, outsourcing responsibility to cloud providers, receiving usability rather than privacy feedback, uneven access to dedicated security staff, ambiguous school AI governance, multi-layer vetting and approval barriers, data-intensive commercial incentives, and inconsistent or weakly enforced regulation. We use the categories as an explanatory structure for the findings below, not as evidence that every EdTech organization exhibits each pattern.

### 3.3 Privacy policy audit

We audited 48 platforms across US K-12 (n=18), US higher education/adult learning (n=14), India K-12/exam preparation (n=11), India higher education (n=2), and AI-native tools (n=3). Platforms were selected for market prominence within each segment using publicly reported usage, app-store rankings, and institutional-adoption data. Segment sizes reflect market maturity and the availability of publicly accessible policy documentation rather than proportional representation. The AI-native group is a functional category rather than a geographic segment; because the India higher-education and AI-native groups are small, we do not treat them as independently representative populations in cross-segment inference. Only documentation reachable within one or two clicks from a platform homepage under headings such as “Privacy,” “Terms of Use,” “Legal,” or “AI/Data Practices” was scored; pages requiring login or buried in support portals were excluded. A zero therefore indicates no accessible disclosure within this scope, not necessarily the absence of an internal commitment. Fragmented or chained policies were noted as an accessibility problem consistent with prior privacy-policy research.

Table 2. Privacy policy audit rubric.

| Dimension | Name | 0 | 1 | 2 |
|---|---|---|---|---|
| C1 | Data collection | No description | Vague categories | Specific types + minimization |
| C2 | Third-party sharing | Not mentioned | Generic partners | Named parties/sub-processors |
| C3 | Children & consent | No engagement | Law cited | Law + practice + dedicated section |
| C4 | AI & automated decisions | No mention | Generic personalization | AI named + training/decisions |
| C5 | Accountability & breach | None | Generic contact/language | Named contact + timeline + standard |

The first author developed the five-dimension rubric from the interview findings and prior EdTech privacy-disclosure literature [9], [10]. The dimensions capture data collection, third-party sharing, children’s consent, AI/automated decision-making, and accountability/breach response, and were designed to distinguish a topic being merely mentioned from operationally meaningful disclosure.

Two coders independently evaluated each platform, assigning a score of 0, 1, or 2 for each dimension.

After discussing interpretive ambiguities, both could revise their scores; adjudicated dimension scores were the mean of revised ratings. Inter-rater reliability was strong overall, with lower agreement concentrated in AI and accountability language, the same areas where disclosure was most ambiguous.

Table 3. Inter-rater reliability results.

| Dimension | % Exact | % +/-1 | Kappa | Interpretation |
|---|---|---|---|---|
| C1 Data collection | 95.8 | 100.0 | 0.864 | Almost perfect |
| C2 Third-party | 93.8 | 97.9 | 0.878 | Almost perfect |
| C3 Children | 91.7 | 97.9 | 0.871 | Almost perfect |
| C4 AI disclosure | 77.1 | 97.9 | 0.654 | Substantial |

| Dimension | % Exact | % +/-1 | Kappa | Interpretation |
|---|---|---|---|---|
| C5 Accountability | 81.2 | 100.0 | 0.645 | Substantial |
| **Mean** | **87.9** | **98.7** | **0.782** | |

Total scores correlated strongly between raters (Pearson r=0.865, p<0.001); all disagreements were by one point.

Several additional practitioners could not participate because of NDAs and organizational compliance restrictions. We therefore do not claim analytical saturation or generalizability from this 12-person sample. Instead, the qualitative component is exploratory and theory-building: we look for converging mechanisms across different roles and organizational contexts, and treat the resulting framework as a set of patterns to test in larger and more representative studies. The 48-platform audit provides complementary structural evidence, but it does not turn the interview sample into a representative one.

# 4 RESULTS

## 4.1 Interview Findings

Interview accounts converged around five connected mechanisms: privacy was repeatedly deferred, responsibility was delegated across actors, weak feedback loops reduced pressure to act, AI made downstream data use harder to see, and uneven organizational capacity shaped who could respond. These themes are presented as an explanatory pattern across this purposive sample, not as estimates of how common each practice is across the EdTech sector.

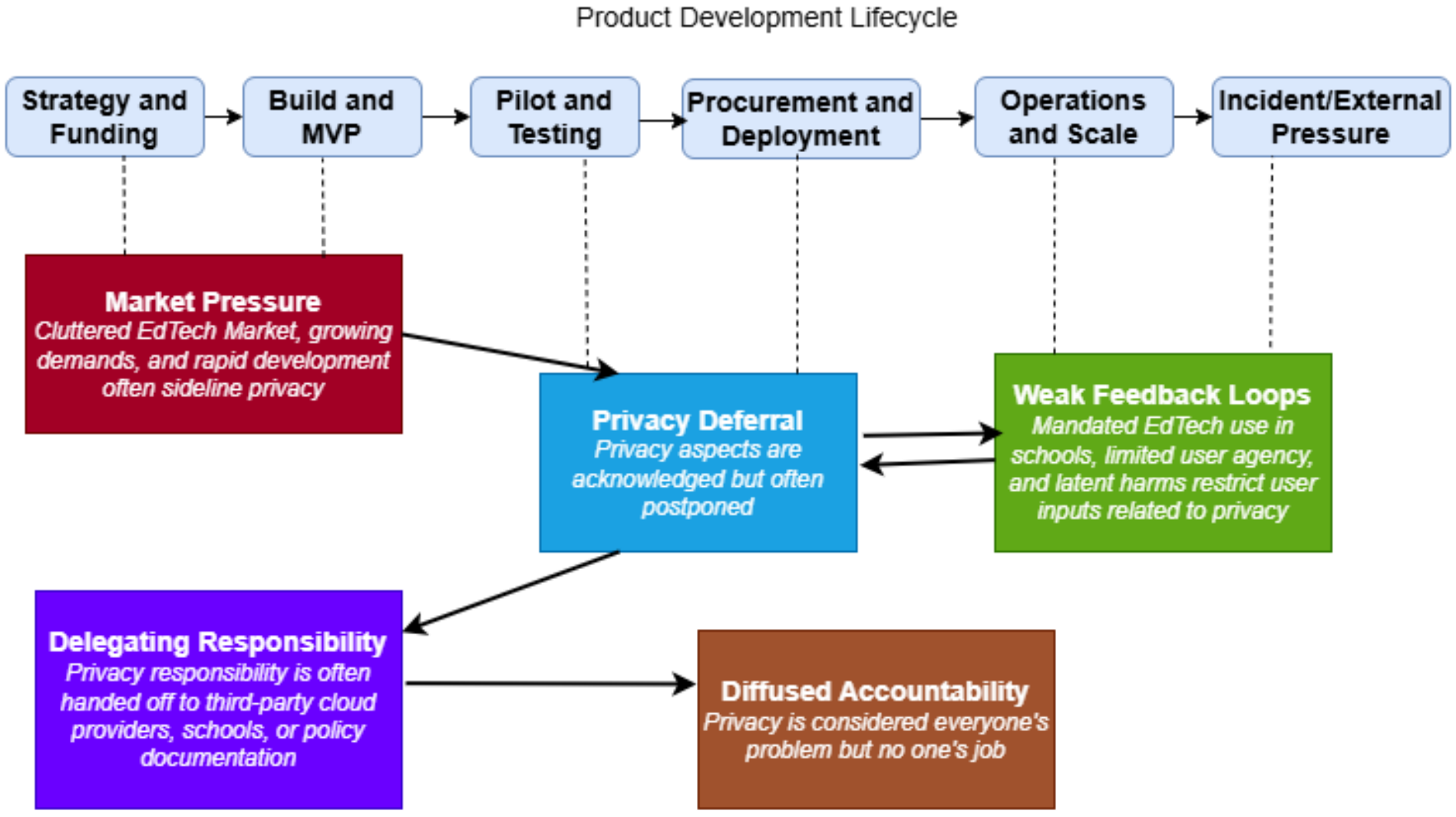

Figure 1. Privacy deferral across the EdTech product lifecycle. Market pressure and delegation stabilize deferral; weak feedback loops sustain it; accountability becomes diffuse.

Participants also distinguished scarcity from strategic deprioritization. In schools, nonprofits, and smaller organizations, privacy capacity was often constrained by staffing, budget, or expertise. In some commercial settings, however, extensive behavioral tracking and personalization were tied directly to the value of the product, creating a different incentive to collect rather than minimize data. AI-related risks surfaced independently across roles even though AI was not the primary interview focus. Participants described inconsistent school capacity to evaluate AI tools, teacher uploads of student information to general-purpose models, and governance that often became reactive only after a problem emerged. These observations position AI as a governance frontier within the broader deferral cycle rather than a separate category of failure.

### Privacy Deferral and Prioritization

Across all 12 interviews, privacy was recognized but consistently deferred. Participants rarely described privacy as unimportant; instead, it was something organizations intended to address after pilots, growth, or core functionality. Participants frequently discussed privacy and security as coupled governance responsibilities; accordingly, the following security-focused account is used as evidence of the broader sequencing through which protective work was postponed. P10 summarized the sequencing directly: *“Security was going to become a second base layer... Once we had our pilots going, then we’d start implementing what was necessary.”* The driver differed by context. Startups faced market-speed pressure, while nonprofits and schools had to justify privacy work against visible program costs. P11 described this as the need to make a “business case” for privacy: teams ask what return an investment will produce and whether it will translate into student outcomes, growth, or future funding.

**Delegation and Diffuse Accountability**

Delegation gave this deferral a structural form. Organizations routinely shifted responsibility to cloud providers, policy documents, compliance functions, or downstream institutions. P4 explained, *“We relied heavily on AWS services for data storage,”* while P12 similarly noted, *“We used Azure’s cloud services to store every piece of data, including user data and the company’s codebase.”* Basic controls such as encryption and two-factor authentication were common, but threat modeling, provenance tracking, and systematic privacy review were more often described as desirable practices that became urgent only after compliance pressure or an incident.

Responsibility also moved downstream to schools. P10 described smaller vendors bypassing district procurement: *“Some companies are going straight to the school and saying, ‘You’re going to give me the data.’ That bypasses the usual process and schools often don’t have the equipment or understanding to manage that layer.”* Each handoff can appear reasonable in isolation: developers rely on cloud defaults, schools rely on vendor documentation, and users rely on institutions. Together, however, these handoffs diffuse ownership until privacy becomes everyone’s responsibility in principle and no one person’s job in practice.

**Weak Feedback Loops**

Weak feedback loops then make the delay sustainable. Participants consistently received usability, engagement, and learning-outcome feedback, but little direct privacy feedback. P6 described a typical process: *“We begin with a sandbox testing phase where a small group evaluates the platform before a full rollout. Direct user feedback on security and privacy features is less common. Most feedback comes in the form of usability ratings.”* P2 similarly observed little end-user pushback, attributing some of the silence to the belief that *“I’m not that important that my data is going to be so valuable.”* Privacy harms are often probabilistic, delayed, and difficult to attribute, so the absence of complaints can be interpreted internally as evidence that no urgent problem exists.

**AI as an Amplifier of Existing Governance Gaps**

AI intensifies this opacity rather than creating a wholly new organizational problem. AI-related concerns emerged across multiple roles, particularly around opaque downstream data use and institutional capacity to evaluate new tools. P7 asked, “How do we use AI to actually teach kids to think deeper or be more evaluative of what comes out?” Yet governance capacity varied sharply. P3 described a school without a designated AI or privacy specialist: “We did not have a designated person for that. We had our principal, who was our go-to tech guy, and then if there was anything that was curriculum-specific, it went through the instructional coaches for that subject.” Participants described reactive governance, including tools being restricted only after problems surfaced, and uncertainty around teachers uploading student data to general-purpose LLMs.

**Unequal Capacity and Commercialization Pressures**

Capacity was unevenly distributed. Larger organizations could maintain dedicated privacy or security staff, whereas smaller organizations assigned those responsibilities alongside other roles. P2 captured the cost tension: *“Maintaining privacy comes at a cost... New organizations might have to forego some aspects of their profit if they don’t collect the data or personalize the tool.”* P8 was more explicit about organizational shortcuts: *“Many organizations take several shortcuts when it comes to securing things because it’s not fun. No one really likes it.”* These accounts distinguish genuine scarcity from strategic deprioritization. That distinction matters because in some commercial settings data collection was not merely an operational byproduct but part of the value proposition. Participants described personalization, aggressive marketing data use, IP tracking, and passive behavioral collection as pressures that can work against minimization. Privacy deferral can therefore arise through different mechanisms: organizations may lack staff, budget, or expertise to act, or they may have incentives to preserve data-intensive practices. Procedural gatekeeping and regulatory ambiguity further shaped when these pressures became visible and who was expected to respond.

**From Themes to the Deferral Cycle**

Figure 1 summarizes the resulting theory. Privacy deferral reproduces itself across the product lifecycle: privacy struggles to enter the business case at strategy and funding stages; cloud delegation can substitute for internal engineering during build; pilots surface usability rather than privacy concerns; procurement can shift responsibility to schools with uneven capacity; AI introduces less visible downstream uses; and incidents finally create reactive attention. The cycle continues because the underlying incentive, ownership, and feedback structures remain unchanged.

### 4.2 Policy Audit Findings

The audit corroborated the interview findings. Data collection disclosure (C1) was strongest, while downstream governance was weaker. Thirty-three percent of platforms (16/48) made no meaningful AI disclosure despite visible AI features. Accountability was similarly shallow: 73% received the middle score, typically a generic email address, vague breach language, or claims of 'industry-standard' security without operational detail.

Table 4. Adjudicated policy scores across 48 platforms.

| Dimension | Mean | SD | Score 0 | Score 1 | Score 2 |
|---|---|---|---|---|---|
| C1 Data collection | 1.81 | 0.38 | 0% | 21% | 79% |
| C2 Third-party sharing | 1.46 | 0.49 | 0% | 56% | 44% |
| C3 Children/consent | 1.05 | 0.87 | 35% | 27% | 38% |
| C4 AI disclosure | 0.90 | 0.79 | 33% | 44% | 23% |
| C5 Accountability | 1.07 | 0.56 | 8% | 73% | 19% |
| **Total (max 10)** | **6.29** | **1.88** | **-** | **-** | **-** |

Segment differences were consistent with external pressure. US K-12 platforms scored highest overall (M=7.47), followed by US higher education/adult learning (M=5.50) and India K-12/exam preparation (M=4.77); differences were significant (Kruskal-Wallis H=20.634, $p<0.001$). K-12 platforms scored significantly higher on children's consent, the dimension most directly governed by COPPA and FERPA, but showed no comparable advantage on AI disclosure or accountability where no equivalent mandate exists. This suggests regulation improves the dimensions it specifically makes enforceable rather than producing generalized privacy maturity.

## 5 DISCUSSION AND IMPLICATIONS

Across both methods, privacy deferral appears self-reinforcing rather than accidental. Privacy work competes with visible product or program outcomes; delegation makes delay feel managed; weak feedback makes harms less salient; and the resulting silence justifies further delay. This extends prior work on later-stage privacy engineering and incentive misalignment [13], [28], [29] by showing how those dynamics operate in a domain where many users cannot exit and where the consequences of a design decision may surface much later.

### 5.1 Why the Deferral Cycle Persists

The cycle persists because each individual decision can appear locally rational. Using cloud defaults is cheaper than building internal privacy engineering, pointing to a policy is easier than operationalizing it, and waiting for complaints is less costly than proactively auditing data flows. P8 captured the temporal logic: *“Someone, oftentimes me, has to step in and say: it’s much better for us to take our vitamins now than to have to take a serious dose of painkillers later.”* Awareness therefore does not automatically produce action. The organizational reward structure favors visible functionality now while the cost of privacy failure is delayed, diffuse, and often borne by students rather than the organization.

The policy audit provides a structural analogue. Platforms score strongly on describing collection in part because disclosure is commonly required by privacy law and procurement expectations; it is also comparatively easier to document than governance practices that require clear internal ownership. Meaningful AI disclosure requires understanding and documenting model and data flows; meaningful accountability requires naming responsible actors, timelines, and standards. The result is a compliance-practice gap: organizations can satisfy visible disclosure expectations without building the mechanisms needed to govern downstream use.

### 5.2 Why EdTech is Different

Privacy deferral is not unique to EdTech. Prior research has shown that software organizations often recognize privacy as important while postponing its implementation until later stages of development, particularly when privacy competes with product functionality, development speed, or other business priorities [13], [14]. What makes EdTech particularly concerning is its weak corrective mechanisms and the vulnerability of children. Students often do not choose their Learning Management System (LMS) or classroom applications, and families may not know which services are in use [15], [16]. Harms such as profiling, secondary use, or altered learning trajectories accumulate quietly and are difficult to trace to a specific decision [1], [3], [21]. This makes user exit, reputational pressure, and ordinary market feedback unreliable correctives.

The K-12 consent result is therefore especially informative. Platforms subject to direct children's privacy obligations performed better on the dimension those rules govern, but the advantage did not generalize to AI disclosure or accountability. External pressure can change behavior, but the effect is targeted: organizations improve where requirements are specific, visible, and enforceable. This supports treating procurement and regulation as governance mechanisms rather than assuming that general privacy awareness will diffuse across practices.

### 5.3 AI as an Amplifier

AI makes existing weaknesses harder to inspect. A parent may understand that a platform stores grades but not that a teacher's spreadsheet upload can transmit student data to a third-party model, or that an automated recommendation can shape educational pathways. One-third of audited platforms made no meaningful AI disclosure on their primary policy pages despite visible AI features. This opacity weakens already-limited feedback loops because users may not know what processing occurred, which actor made a decision, or where to direct a concern.

### 5.4 Practical Implications

For developers and firms, privacy should become a release criterion rather than a documentation exercise. A practical stage-gate could require a one-page data-flow summary completed by the feature owner before student-facing deployment: what data is collected, which third parties receive it, whether it is retained or used for model training, what automated decisions occur, and who owns incident response. Just-in-time notices are particularly important for AI features; when an educator uploads a roster or spreadsheet, the interface should make third-party transfer and model-training use visible at the point of action. Clear internal ownership can also prevent accountability from dissolving across cloud, legal, and compliance teams [28], [29].

For schools, administrators, and policymakers, procurement should demand operational evidence rather than accepting the existence of a privacy policy as sufficient. Vendors should be required to specify collected fields, subprocessors, children's consent practices, AI uses, retention or training practices, and breach-notification timelines in plain language. Institutions can use the five audit dimensions in Table 2 as a lightweight comparison rubric. The results also suggest that regulation should be specific: the K-12 consent advantage shows that enforceable obligations can raise disclosure quality on the dimension they directly govern.

For educators, the findings argue for structured support rather than placing the burden on individual judgment. Teachers often become the final adoption point for tools but may lack time, legal expertise, or visibility into data flows. Training should therefore focus on concrete decisions: recognizing when student information is being transferred, checking whether AI use is disclosed, and knowing which institutional channel owns approval. Privacy literacy and student-facing data education can strengthen feedback over time, but literacy cannot substitute for organizational accountability; students and teachers should not be expected to compensate for opaque vendor practices.

For researchers, the findings caution against treating privacy as a single maturity score. The audit dimensions capture distinct governance choices: a platform that clearly describes collection is not necessarily more transparent about AI or more accountable for breaches. Future work should track these dimensions longitudinally, especially as India's DPDPA enforcement matures, and examine whether procurement requirements or AI-specific rules create spillover beyond the dimensions they directly regulate. Larger studies should also include commercial platform decision-makers and regulators who were difficult to access in the present sample.

## 6 LIMITATIONS

The interview sample is small (n=12), primarily US-based, and shaped by NDA and compliance restrictions; large commercial platforms and regulators are not directly represented. Some participants were closer to implementation than strategic decision-making, which limits visibility but also reveals how governance is experienced operationally. The policy audit is a snapshot of publicly accessible documentation, not proof of internal behavior, and the India segment is smaller than its market importance. DPDPA enforcement was also still developing during the study, so cross-country findings should be interpreted cautiously. The US-India score differences should therefore be read as patterns within this purposively assembled audit sample rather than estimates of country-level privacy maturity; unequal segment sizes, platform mix, and different stages of regulatory enforcement limit direct national comparison.

## 7 CONCLUSIONS

Privacy in EdTech is not simply overlooked; it is repeatedly deferred. Across 12 interviews and 48 policy audits, organizations recognized privacy risks but postponed action, delegated responsibility, and received too little feedback to make those risks urgent. Policies mirrored this structure: data collection was usually described, while AI governance and meaningful accountability were much weaker. Because students often cannot choose or leave the systems that collect their data, the usual corrective mechanisms of consumer markets are limited. Closing the gap therefore requires institutional procurement standards, clearer internal ownership, and enforceable oversight that makes privacy a condition of deployment rather than a promise to revisit later.

# ACKNOWLEDGEMENTS

We sincerely thank all professionals who participated in this study and shared their insights despite time and organizational constraints. We also thank Julia Jose for her assistance with the privacy policy audit.